\documentclass[a4paper,11pt]{article}

\pdfoutput=1
\usepackage{jcappub}

\usepackage{xcolor} 
\usepackage{hyperref}
\usepackage{listings}
\usepackage{mathtools}
\usepackage{amsmath}
\usepackage{amsfonts}
\usepackage{amssymb}
\usepackage{bbold}
\usepackage{epsfig}
\usepackage{graphicx}
\usepackage{bm}
\usepackage{array}
\usepackage{hyperref}
\usepackage{listings}
\usepackage{color}
\usepackage{float}
\usepackage[normalem]{ulem}
\usepackage{placeins}
\usepackage{bbold}

\newcommand{\exclude}[1]{}

\usepackage{tikz,xcolor,hyperref}

\definecolor{lime}{HTML}{A6CE39}
\DeclareRobustCommand{\orcidicon}{\hspace{-1mm}
 \begin{tikzpicture}
 \draw[lime, fill=lime] (0,0) 
 circle [radius=0.16] 
 node[white] {{\fontfamily{qag}\selectfont \tiny \,ID}};
 \draw[white, fill=white] (-0.0525,0.095) 
 circle [radius=0.007];
 \end{tikzpicture}
 \hspace{-3mm}
}

\foreach \x in {A, ..., Z}{\expandafter\xdef\csname orcid\x\endcsname{\noexpand\href{https://orcid.org/\csname orcidauthor\x\endcsname}
 {\noexpand\orcidicon}}
}

\title{On the length scale of collective neutrino oscillations}

\author[a]{Shashank Shalgar\orcidB{}}

\affiliation[a]{Niels Bohr International Academy \& DARK, Niels Bohr Institute,\\University of Copenhagen, Blegdamsvej 17, 2100 Copenhagen, Denmark}

\emailAdd{shashank.shalgar@nbi.ku.dk}

\abstract{ 

In this paper, I present a discussion on the length scale of collective neutrino oscillations. There is a popular myth in the field that the length scale of collective neutrino oscillation is related to the strength of self-interaction potential; this is a result of confusion between the length scale and time scale. As a consequence of this myth, it is believed that the convergence of numerical simulation of quantum kinetic equations requires a spatial resolution (radial bin size) that is equal to the inverse of the self-interaction potential. I try to debunk this myth in this paper.

This document will NOT be submitted to any journal; this document has been prepared for submission to arXiv only.

}

\begin{document}
\maketitle

\section{Introduction}

In dense astrophysical systems such as core-collapse supernovae and neutron star mergers, neutrinos can experience significant coherent forward scattering (refraction) due to other neutrinos. This leads to a modification of neutrino flavor evolution like the modification neutrino flavor evolution due to the matter effect~\cite{1978PhRvD..17.2369W, Mikheev:1986if}. However, unlike the matter effect, the modification of neutrino flavor evolution due to coherent forward scattering of neutrinos with other neutrinos (neutrino self-interactions) is nonlinear in nature~\cite{Pantaleone:1992eq, Sigl:1992fn}.

Although neutrino self-interactions have been known since the 1990s, the interest in the field grew dramatically with the realization that these effects can be important in the context of core-collapse supernovae~\cite{Duan:2005cp,Duan:2006an, Duan:2006jv,Duan:2007bt, Fogli:2007bk,Duan:2010bg}. This realization came largely by studying neutrino flavor evolution in the context of the `neutrino-bulb model', which made several simplifying assumptions. The most significant assumptions used in the neutrino-bulb model were that of spherical symmetry and instantaneous decoupling of neutrinos at a fixed radius that is the same for all flavors. Through numerical simulations, it was shown that the neutrino flavor evolution for various momentum modes is correlated. Consequently, neutrino flavor evolution in the presence of self-interactions is used interchangeably with `collective neutrino oscillations' Neutrino self-interactions in this model lead to flavor evolution only when the strength of neutrino self-interactions is comparable to the vacuum frequency. This leads to neutrino flavor evolution on time scales that are not too short, and hence called `slow collective oscillations.'

In the presence of a nontrivial angular distribution of neutrinos, it is possible for there to be a crossing between the angular distribution of electron lepton number (ELN), which can lead to neutrino flavor evolution even when the strength of neutrino self-interaction approaches infinity (or equivalently, in the limit of vanishing vacuum frequency). This category of neutrino flavor evolution is generally called `fast collective oscillations,' as they can occur on a time scale that is given by the inverse of the strength of the neutrino self-interactions (generally denoted by $\mu$)~\cite{Sawyer:2005jk,Sawyer:2008zs,Sawyer:2015dsa,Chakraborty:2016yeg,Tamborra:2020cul}. 

Modeling neutrino flavor evolution in the interior of a core-collapse supernova is not easy due to the problem's nonlinearity. Significant numerical resources are needed to model even a simple system. To make matters worse one has to take into account the effect of advection and collisions to gain insight into the magnitude of flavor evolution. Such calculations have been possible only recently but have also brought with them a controversy regarding the length scale of the problem, which I address in this paper. 

The time scale associated with collective neutrino oscillations is given by $\mu^{-1}$, where $\mu$ is the strength of the self-interactions as explained earlier. Of course, this holds only when the significant flavor evolution does happen in the presence of neutrino self-interactions. For a flavor-stable system, the time scale of evolution is determined by the time scale over which the collision rates change which is very slow in comparison. It is sometimes argued that since neutrinos travel at the speed of light, $c$, the length scale of the system is given by $c\mu^{-1}$. This is also extended to claim that in numerical simulations the spatial bins should have a size of $c\mu^{-1}$ (a few cm) or smaller. I argue against that in this paper.

This paper is organized as follows: In Sec.~\ref{sechomo}, I will discuss flavor evolution in a homogeneous gas of neutrinos and inhomogeneous neutrino gas with periodic boundary conditions in Sec.~\ref{perbox}. In Sec.~\ref{qkes}, I introduce quantum kinetic equations in spherical geometry; in Sec.~\ref{spsym} I will discuss the length scale of the problem and conclude in Sec.~\ref{conclusion}.

\section{Homogeneous neutrino gas}
\label{sechomo}

Although this paper is about length scale of collective neutrino oscillations let us begin by considering a homogeneous system of neutrinos undergoing fast flavor evolution. This will give us some insights into the convergence with respect to the number of angle bins and what happens if insufficient number of angle bins are used.

For the case of homogeneous neutrinos in the single energy, two flavor approximation, and no collision terms included, the equations of motion are given by,
\begin{eqnarray}
\label{eomhom1}
i\frac{\partial \rho(\cos\theta,t)}{\partial t} &=& [H(\cos\theta,t),\rho(\cos\theta,t)]\\
\label{eomhom2}
i\frac{\partial \bar{\rho}(\cos\theta,t)}{\partial t} &=& [\bar{H}(\cos\theta,t),\bar{\rho}(\cos\theta,t)]\ .
\end{eqnarray}
Here, $\rho(\cos\theta,t)$ and $\bar{\rho}(\cos\theta,t)$ are $2 \times 2$ density matrices that encode the flavor state of neutrinos and antineutrinos, respectively. They can be written explicitly in the following form,
\begin{eqnarray}
\rho(\cos\theta,t) = 
\begin{pmatrix}
\rho_{ee}(\cos\theta,t) & \rho_{ex}(\cos\theta,t) \\
\rho_{ex}^{*}(\cos\theta,t) & \rho_{xx}(\cos\theta,t)
\end{pmatrix} 
\quad \quad 
\bar{\rho}(\cos\theta,t) = 
\begin{pmatrix}
\bar{\rho}_{ee}(\cos\theta,t) & \bar{\rho}_{ex}(\cos\theta,t) \\
\bar{\rho}_{ex}^{*}(\cos\theta,t) & \bar{\rho}_{xx}(\cos\theta,t)
\end{pmatrix} 
\end{eqnarray}
The Hamiltonians $H(\cos\theta)$ and $\bar{H}(\cos\theta)$ are for neutrinos and anti-neutrinos, respectively, and $[\ldots,\ldots]$ represents a commutator.
Ignoring the matter term, the Hamiltonians are given by,
\begin{eqnarray}
H(\cos\theta) &=& H_{\textrm{vac}} + H_{\nu\nu} \\
\bar{H}(\cos\theta) &=& -H_{\textrm{vac}} + H_{\nu\nu} \\
H_{\textrm{vac}} &=& \frac{\omega_{\textrm{vac}}}{2}
\begin{pmatrix}
-\cos 2\vartheta_{\textrm{V}} & \sin 2\vartheta_{\textrm{V}} \\
\sin 2\vartheta_{\textrm{V}} & \cos 2 \vartheta_{\textrm{V}}
\end{pmatrix} \\
H_{\nu\nu}(\cos\theta) &=& \mu \int_{-1}^{1} [\rho(\cos\theta^{\prime}) - \bar{\rho}(\cos\theta^{\prime})] (1-\cos\theta^{\prime} \cos\theta) d\cos\theta^{\prime}\ .
\end{eqnarray}
Let us consider this system with the following initial conditions and parameters,
\begin{eqnarray}
\label{rhoeeini}
\rho_{ee}(\cos\theta) &=& 0.5\\
\label{rhoeebarini}
\bar{\rho}_{ee}(\cos\theta) &=& 0.47 + 0.05 \exp\left(-(1-\cos\theta)^{2}\right)\\
\rho_{xx} &=& \bar{\rho}_{xx} = 0\\
\mu &=& 100~\textrm{km}^{-1} c\\
\omega_{\textrm{vac}} &=& 0.001~\textrm{km}^{-1} c\\
\theta_{\textrm{V}} &=& 10^{-3}\ .
\end{eqnarray}

Since the neutrino gas is homogeneous, by definition, the length scale associated with the problem is infinite. This simple scenario demonstrates that the time scale is not related to the length scale by a simple multiplication by $c$. 

I have chosen the values of $\mu$ and $\omega_{\textrm{vac}}$ such that the ratio $\mu/\omega_{\textrm{vac}}$ is very big. This ensures the absence of rapid oscillations, making it easier to read the plots. However, even for a smaller value of the ratio $\mu/\omega_{\textrm{vac}}$ the conclusions remain unchanged. 

\begin{figure}
\includegraphics[width=0.49\textwidth]{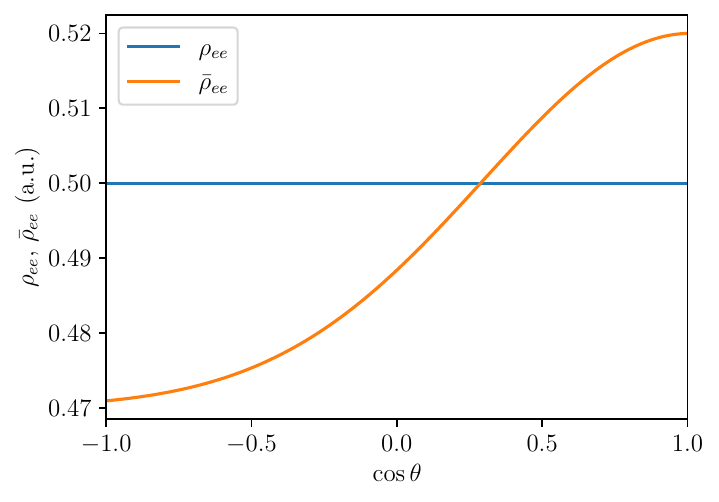}
\includegraphics[width=0.49\textwidth]{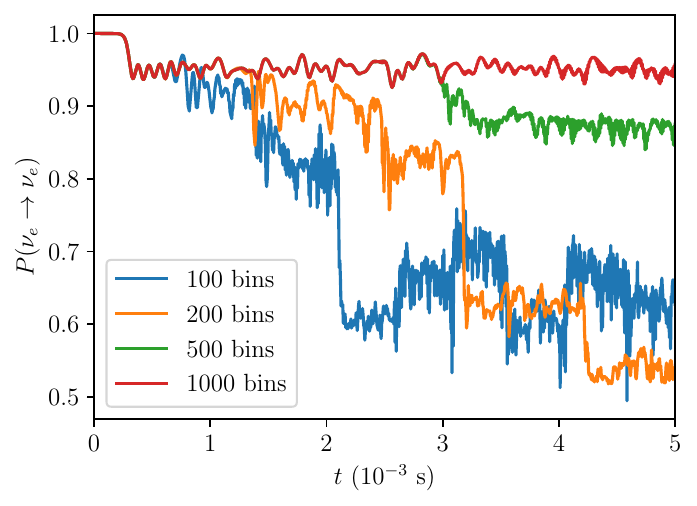}
\caption{{\it Left:} Initial angular distribution of $\rho_{ee}$ and $\bar{\rho}_{ee}$ used in the simulation of homogeneous gas in this section (see eqs.~\ref{rhoeeini} and \ref{rhoeebarini}). {\it Right:} Evolution of angle averaged survival probability as a function of time with various numbers of angle bins. It can be seen that the simulation with 100 bins (blue line) gives correct results up to about $t \approx 0.6 \times 10^{-3}$~s and then deviated from the results obtained using finer resolution. For the simulation with 200 angle bins (orange line), the results are correct for a longer period of time. The result for 500 angle bins (green line) is identical to the result obtained using 1000 angle bins (red line) upto the simulation time of $\approx 3\times10^{-3}$~s. If the simulation is carried out for a longer period of time, this trend persists, and more angle bins are required as the simulation time increases.}
\label{homogeneous}
\end{figure}

In Fig.~\ref{homogeneous}, I show the initial angular distributions and the temporal evolution of the angle-integrated flavor content. It can be seen that the longer the system is evolved the more number of angle bins are required for convergence.
From Fig.~\ref{homogeneous}, it is clear that there isn't a minimum angular scale associated with the problem in general; it depends on the time for which the system has been evolved. The only exception is the case of $\omega_{\textrm{vac}} = 0$, with a perturbation to trigger flavor evolution. In this case, the flavor evolution is bipolar, and there is no cascade to smaller angular scales.

Even in the absence of a sufficient number of angle bins, the survival probability neither exceeds 1 nor goes below 0. This is not just true for the angle averaged transition probability but also for each angle bin. This is because the lack of a sufficient number of angle bins causes the Hamiltonian to be inaccurate but the Hamiltonian is still Hermitian. In the next section, we relax the condition of homogeneity and investigate the changes.

\section{Neutrino gas with periodic boundary conditions}
\label{perbox}

Now, let us make things slightly more complicated by adding some spatial structure to the problem and trying to understand whether a spatial bin size of $c\mu^{-1}$ would be required for numerical simulations to be reliable. 

Let us consider a box with sides $L$ and periodic boundary conditions for the sake of simplicity. We can try to estimate the length scale of the problem and, hence, the resolution required for a numerical simulation. Also, let us assume that the neutrino gas remains completely homogeneous in the $x$ and $y$ directions, and any possible inhomogeneity can only arise in the $z$ direction. Let us begin with a perturbation of the form $\sin(2\pi z/L)$ in the diagonal component of the density matrices. 

In that case, the eqs.~\ref{eomhom1} and \ref{eomhom2} are modified to the ones below,
\begin{eqnarray}
\label{eominhom1}
i\frac{\partial \rho(\cos\theta,z,t)}{\partial t} &=&  -i\cos\theta\frac{\partial \rho(\cos\theta,z,t)}{\partial z} + [H(\cos\theta,z,t),\rho(\cos\theta,z,t)]\\
\label{eominhom2}
i\frac{\partial \bar{\rho}(\cos\theta,z,t)}{\partial t} &=& -i\cos\theta\frac{\partial \bar{\rho}(\cos\theta,z,t)}{\partial z} + [\bar{H}(\cos\theta,z,t),\bar{\rho}(\cos\theta,z,t)]\ .
\end{eqnarray}
The derivative on the right-hand side can be calculated by the central difference method. 
The Hamiltonian, initial conditions and the parameters are kept the same as the ones in the previous section except for eq.~\ref{rhoeeini} which is modified as below:
\begin{eqnarray}
\rho_{ee} = 0.5 + 10^{-3} \sin\left(\frac{2\pi z}{L}\right) \ ,
\end{eqnarray}
where, I have used $L$ = 10 km.

\begin{figure}
\includegraphics[width=0.49\textwidth]{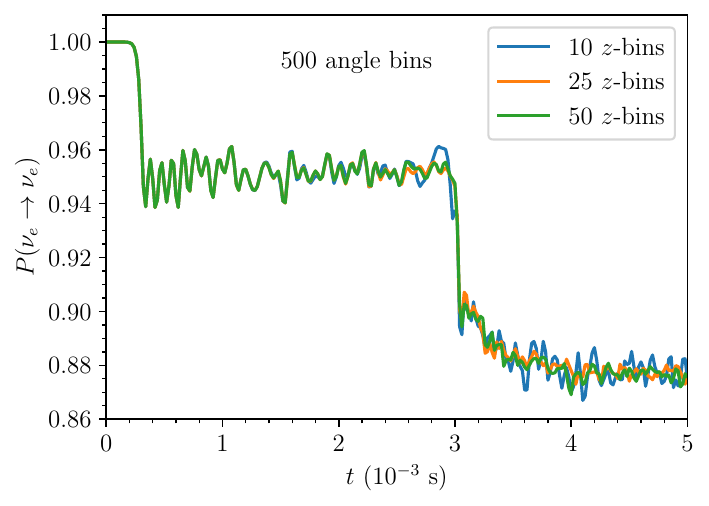}
\includegraphics[width=0.49\textwidth]{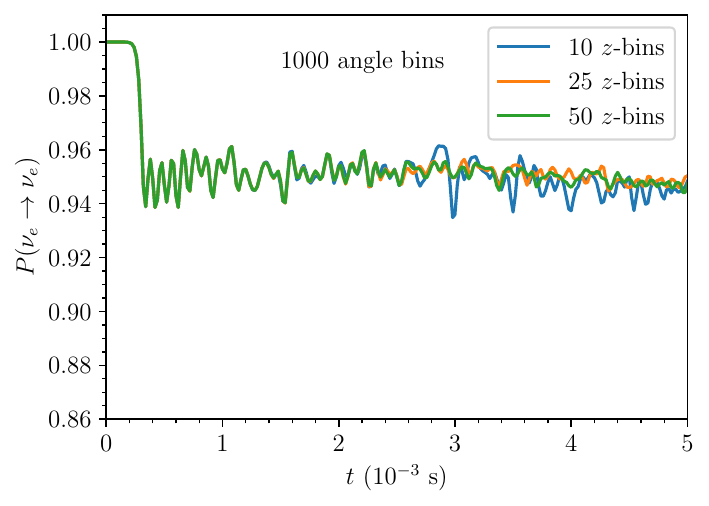}
\caption{{\it Left:} Angle and spatially averaged survival probability of $\nu_{e}$ with 500 $\cos\theta$ bins. The blue, orange, and green lines show the survival probability as a function of time with 10, 25, and 50 spatial bins. It is clear that the feature at around $3\times 10^{-3}$~s is due to an insufficient number of angle bins. {\it Right:} The same plot as the left panel but with 1000 angle bins. The divergence between the three lines indicated the time at which the number of spatial bins is insufficient for numerical convergence and the magnitude of error due to lack of convergence.}
\label{inhom}
\end{figure}

Figure~\ref{inhom} shows the angle and $z$-averaged survival probability for the case described in the setup mentioned above. In the left panel of Fig.~\ref{inhom}, which uses 500 $\cos\theta$ bins, one can see a sharp feature at around $t \approx 3 \times 10^{-3}$~s which is due to not having a sufficient number of angle bins. This feature is missing in the right panel of Fig.~\ref{inhom} due to a sufficient number of angle bins. The time for which the system in Fig.~\ref{inhom} has been evolved is much larger $L/c$, which means that if the system did not satisfy periodic boundary conditions, the spatial structure would not develop significantly within the time taken for neutrinos to transit the box. Within that time range, even 10 bins in the $z$-direction give sufficiently good results. This is not that surprising. As seen in eqs.~\ref{eominhom1} and \ref{eominhom2}, the term involving the commutator is responsible for the cascade to a smaller and smaller angular scale. It is nonlinear and typically larger than the derivative term. So, the development of spatial structure is much more well-behaved. For a periodic box like the one described by eqs.~\ref{eominhom1} and \ref{eominhom2}, it is thus natural that the simulation will run into problems due to not having sufficient angle bins rather than due to the size of the spatial bin. 
There are a few exceptions to this: if the vacuum frequency is set to zero and fast flavor conversions are triggered by a perturbation in the off-diagonal components of the density matrix then there is no cascade to smaller angular scales as mentioned in the previous section, which implies that the number of spatial bins will determine the convergence of the numerical simulations. Also, if the derivative term, $d\rho/dz$, is large initially by choice, then the number of spatial bins becomes more important than the number of angle bins. There are a few papers that use very small size of spatial bins and the perturbation in each spatial bin is assumed to be independent; this will naturally lead to a very large value of $d\rho/dz$. It is equivalent to putting a small length scale by hand instead of a small length scale arising in the system. Lastly, I would like to mention that the system used to produce results in Fig.~\ref{inhom} has not been cherry-picked to make a point. This is what should be expected from any reasonable configuration.
It is clear from comparing the left and right panels of Fig.~\ref{inhom} that the error arising from using an insufficient number of spatial bins is not as serious as the one arising from using an insufficient number of angle bins.

The numerical example shown in Fig.~\ref{inhom} demonstrates that even in a toy model with periodic boundary conditions, it is difficult to justify the need for spatial bins of the size $c\mu^{-1}$. A small spatial bin size is required for convergence in very contrived scenarios. 
It is clear that initially the length scale of the system is $L$, and so the required spatial bin size should be a few orders of magnitude smaller than $L$ ($L/10$ or $L/100$ depending on the required accuracy). As the system evolves, the required bin size should change with time. As long as the system is in the linear regime ($|\rho_{ex}| \ll |\rho_{ee}-\rho_{xx}|$), the Fourier modes are decoupled from each other, and the required size of the spatial bins remains unchanged~\cite{Duan:2014gfa, Abbar:2015mca}. However, as the system enters the nonlinear regime, the Fourier modes are coupled to their neighbors; the structure starts developing at smaller and smaller scales. There have been previous numerical studies that have shown that this leads to a power-law like distribution of Fourier modes, which eventually leads to all Fourier modes having amplitudes within the same order of magnitude~\cite{Richers:2022bkd, Cornelius:2023eop}. This development is similar to the cascade in a turbulent system of fluid dynamics with a large Reynolds number. However, unlike a turbulent system which exhibits a cut-off in the power spectrum due to dissipative processes, there is no cut-off in the neutrino system; the cascade continues to smaller and smaller length scales~\footnote{It should be noted that using the finite-difference method for spatial derivative introduces a fake numerical viscosity but the magnitude of that viscosity depends on the spatial resolution and it is not physical.}. Presumably, there is a scale at which the mean-field formalism breaks down, but it most likely happens at a very small length scale. 

Now coming back to the question of what is the required size of the spatial bins. The required size of the spatial bins depends on the time for which the system is evolved. As the time for which the system has evolved increases, the required size of the spatial bins decreases.

This is similar to the cascade of angular distribution to smaller and smaller angular scales. If we consider fast flavor evolution in a homogeneous gas of neutrinos, the required number of angle bins increases with the time for which the system is evolved. There is no point as far as we know after which the required number of angle bins plateaus~\cite{Johns:2020qsk}. The same thing happens with the number of spatial bins, which is not surprising since the angular modes and spatial modes are coupled. As the time for which the system is evolved increases the number of required angle bins and spatial bins increases in the absence of homogeneity. There is no fixed length scale associated with the problem. 

In order to understand the number of angle and spatial bins required for numerical simulations to be reliable, let us ask the counter question: What happens if the number of angle bins or spatial bins is not sufficient? The answer differs in the case of angle bins and spatial bins. If the number of angle bins used is not sufficient, the self-interaction Hamiltonian can be over or underestimated, but the flavor evolution due to the self-interaction term remains unitary, as discussed in the previous section. The flavor transition probability remains between 0 and 1 despite the lack of a sufficient number of angle bins. However, this cannot be used as a test of convergence. Even if an insufficient number of spatial bins is used, the results are approximately correct, while the same is not true with regard to the number of angle bins as seen in Fig.~\ref{inhom}. Also, it is difficult to see how one can expect a structure at the scale of $c\mu^{-1}$ without seeing a structure at a larger scale that is significant enough to affect the overall outcome. If the number of spatial bins is not sufficient, the system can lose unitarity at a numerical level, and the flavor transition probability can go outside the range of 0 and 1. Of course, the same can happen if the time step size used in the simulation is larger than it should be but we can operate under the assumption that adaptive step size or extremely small time step is used in the simulation.

The origin of the misconception that the length scale associated with collective neutrino oscillations is $c\mu^{-1}$ remains unclear, but it is entertaining to investigate what the consequences would be if it were true. If we perform a Fourier decomposition of the number density along a particular direction, would there be a cut-off at $c\mu^{-1}$ above which all the Fourier modes will have zero amplitude? Or will there be a peak at Fourier modes corresponding to the length scale $c\mu^{-1}$? It is difficult to come up with a mechanism by which this would happen in a periodic box with a neutrino gas undergoing collective neutrino oscillations.

Thus, it can be said that even for a system with periodic boundary conditions, there is no length scale associated with the problem. The length scale and the required resolution depend on the system's initial configuration (the magnitude of initial inhomogeneity) and the time for which the system has been numerically evolved. 
A system with periodic boundary conditions is unrealistic as it involves teleportation of neutrino from $z=0$ to $z=L$ and vice-versa. It is much better to thus focus on a system of QKEs without periodic boundary conditions.

\section{Quantum Kinetic Equations}
\label{qkes}

In this section, I introduce the Quantum Kinetic equations (QKEs) that are used to study neutrino flavor evolution in the mean-field approximation in spherical geometry. These equations are a generalization of the equations introduced in the previous two sections. For the sake of simplicity, I will restrict the analysis to spherical geometry. The imposition of spherical symmetry implies that the system can be described in terms of three independent variables, the radial coordinate, $r$, the polar angle, $\theta$, the energy, $E$, and time, $t$. The density matrices, $\rho(r,\cos\theta,E,t)$ and $\bar{\rho}(r,\cos\theta,E,t)$ now have additional dependence which did not exist earlier. The flavor content of the neutrinos in the interior of a core-collapse supernova is given by the following equations:
\begin{eqnarray}
\label{eom1}
i\left(\frac{\partial}{\partial t} + \vec{v} \cdot \vec{\nabla}\right)\rho(r,\cos\theta,E,t) &=& [H(r,\cos\theta,E,t),\rho(r,\cos\theta,E,t)] + i\mathcal{C}[\rho] \ ,\\
\label{eom2}
i\left(\frac{\partial}{\partial t} + \vec{v} \cdot \vec{\nabla}\right)\bar{\rho}(r,\cos\theta,E,t) &=& [\bar{H}(r,\cos\theta,E,t),\bar{\rho}(r,\cos\theta,E,t)] + i\bar{\mathcal{C}}[\bar{\rho}]\ .
\end{eqnarray}
Here, $\vec{\nabla}$ represent the gradient in spherical coordinates, and the advective term can be written as,
\begin{eqnarray}
\label{graddef}
\vec{v} \cdot \vec{\nabla} = \cos\theta \frac{\partial}{\partial r} + \frac{\sin^{2}\theta}{r} \frac{\partial}{\partial \cos\theta}\ .
\end{eqnarray}
$\mathcal{C}$ ($\bar{\mathcal{C}}$) schematically represent the collision terms for neutrinos (antineutrinos) that will be discussed in detail later.

The Hamiltonians that govern the neutrino flavor evolution consist of three terms, the vacuum term, the matter term, and the self-interaction term, 
\begin{eqnarray}
H(r,\cos\theta,E,t) &=& H_{\textrm{vac}}(E) +H_{\mathrm{mat}}+ H_{\nu\nu}(r,\cos\theta, t)\\
\bar{H}(r,\cos\theta,E,t) &=& -H_{\textrm{vac}}(E) +H_{\mathrm{mat}}+ H_{\nu\nu}(r,\cos\theta,t)\ ,
\end{eqnarray}
where,
\begin{eqnarray}
H_{\textrm{vac}}(E) &=& \frac{\omega_{\textrm{vac}}}{2}
\begin{pmatrix}
-\cos 2\vartheta_{\textrm{V}} & \sin 2\vartheta_{\textrm{V}} \\
\sin 2\vartheta_{\textrm{V}} & \cos 2 \vartheta_{\textrm{V}}
\end{pmatrix}\ ,\\
H_{\nu\nu}(r,\cos\theta,t) &=&  \sqrt{2} G_{\textrm{F}}  \int_{-1}^{1}\int_{0}^{\infty}\left[\rho(r,\cos \theta^{\prime},E,t) - \bar{\rho}(r,\cos \theta^{\prime},E,t)\right] \\
& \times &
(1-\cos\theta \cos\theta^{\prime}) dE ~d\cos\theta^{\prime}\ .
\end{eqnarray}
Here, $\omega_{\textrm{vac}} = {\Delta m^{2}}/{2 E}$, $\theta_{\mathrm{V}}$ is the vacuum mixing angle, and $G_{\textrm{F}}$ is the Fermi constant.
The matter Hamiltonian has the overall effect of suppressing the effective mixing angle and can be ignored as long as the vacuum mixing angle is replaced by a smaller quantity and we ignore it in this paper.

The collision terms $\mathcal{C}$ and $\bar{\mathcal{C}}$ consist of three components corresponding to emission, absorption, and direction-changing terms:
\begin{eqnarray}
\label{coll1}
\mathcal{C}[\rho] &=& \mathcal{C}_{\textrm{emit}}(r,E) - \mathcal{C}_{\textrm{absorb}}(r,E) \odot \rho(r,\cos\theta,E) 
+ \cos\theta\ \mathcal{C}_{\textrm{ani}}(r,E) \int d\cos\theta^{\prime} \cos\theta^{\prime} \rho(r,\cos\theta^{\prime},E)
 \nonumber \\
 &+& \frac{\mathcal{C}_{\textrm{dir-ch}}(r,E)}{2}\int d \cos\theta^\prime \left[-\rho(r,\cos\theta,E)+\rho(r,\cos\theta^{\prime},E)\right] \ ,
\end{eqnarray}
and a similar expression for antineutrinos.
The direction-changing term has been divided into the last two terms that take into account the isotropic and anisotropic components. As the name suggests, it conserves the number of neutrinos, unlike the emission and absorption terms.

In the interior of a supernova, the time scale over which the collision terms $\mathcal{C}$ and $\bar{\mathcal{C}}$ change is very slow compared to all other time scales in eqs.~\ref{eom1} and \ref{eom2}. Also, unlike in the previous section, one does not have to introduce spatially inhomogeneous perturbations by hand. The inhomogeneous perturbation of the problem is determined by the value of the collision terms and their gradients, which typically involve a scale of $\mathcal{O}(1) - \mathcal{O}(10)$~km. The problem of figuring out the flavor content thus reduces to finding the quasi-steady state configuration of eqs.~\ref{eom1} and \ref{eom2} for collision terms at given times. 

This can be done by starting with any configuration and evolving eqs.~\ref{eom1} and \ref{eom2} until the diagonal components of density matrices stop evolving to a reasonable approximation. There are no agreed-upon criteria for how much variation would constitute a quasi-steady state but I will leave that question aside as far as this paper is concerned. The reason why this is called the quasi-steady state and not the steady state is because the off-diagonal components never stop evolving. 
It should be noted that we are actually not interested in the time evolution of the flavor states but only in the quasi-steady state configuration~\cite{Shalgar:2022rjj, Shalgar:2022lvv}. 

\section{Length scale of QKEs in a spherically symmetric supernova}
\label{spsym}

The main question that I wish to address in this paper is regarding the gradient term in the form of the $\vec{v}\cdot\vec{\nabla}$ in eqs.~\ref{eom1} and \ref{eom2}. This term has to be calculated using the central difference method as in the previous section or a higher order stencil. For the sake of concreteness, let us assume that the gradient is calculated using the central difference method. It should be noted that the gradient involves a spatial derivative as well as a derivative with respect to the angular coordinates as seen in eq.~\ref{graddef}. The difference between the two is that one derivative is with respect to a dimensionful quantity while the other is with respect to a dimensionless quantity. It is thus tempting to claim that an arbitrary dimensionful quantity appearing in the problem determines the size of the radial bin required for convergence. The same temptation applied to the angular part would appear to be an unreasonable proposition. Fortunately, this proposition has never been made to my knowledge; however, as I write this statement, I fear that there will be a race to claim that the size of the angular bin required for convergence should be the same as the matter suppressed mixing angle. 

In the rest of this section, I will elaborate on why it is not justified to assume that the length scale of the problem should be assumed to be related to the time scale in a trivial manner as is often assumed. The simplest way to demonstrate this is to consider the problem of QKEs in a spherical system with no self-interactions; only the vacuum term along with advection and collisions. For the sake of simplicity I will demonstrate this in the single energy approximation. This allows for a simulation with a very large number of spatial bins which the readers can try out for themselves.

\begin{figure}
\includegraphics[width=0.49\textwidth]{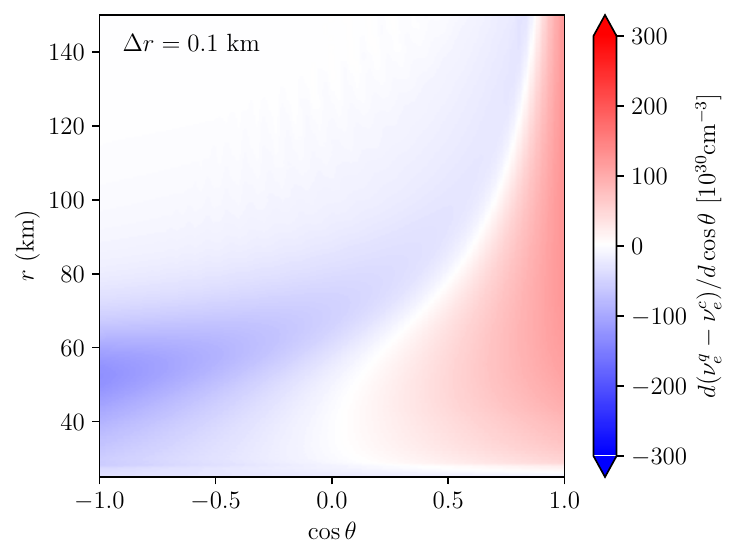}
\includegraphics[width=0.49\textwidth]{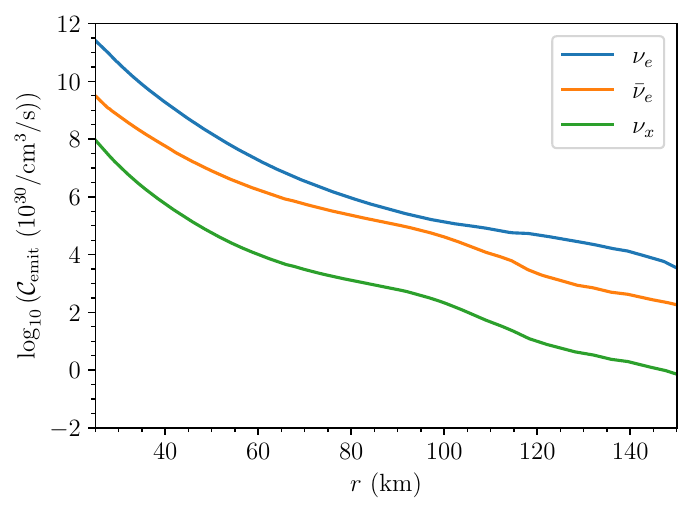}
\includegraphics[width=0.49\textwidth]{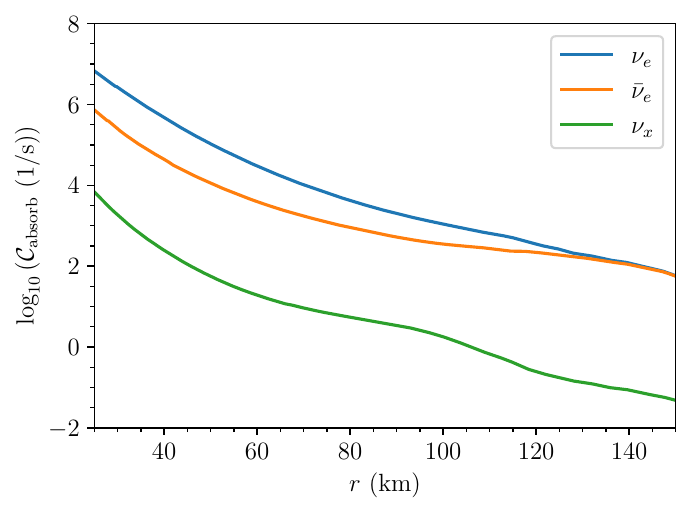}
\includegraphics[width=0.49\textwidth]{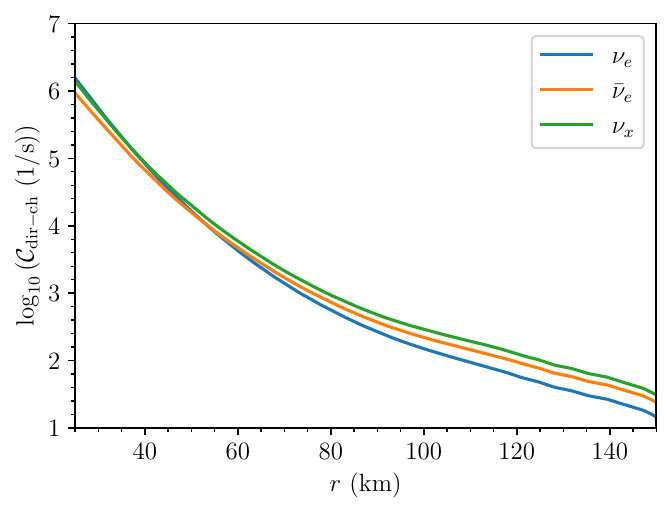}
\caption{{\it Top-left:} Heatmap of change in the number density of $\nu_{e}$ due to vacuum oscillations. The mixing angle, $\theta_{\textrm{V}}$ is assumed to be $\pi/4$ and $c\omega_{\textrm{vac}}^{-1} = 0.1$~km. One would naively expect the length scale of the problem to be 0.1 km, but as seen in the heatmap there is no structure that can be said to be of size 0.1 km. The radial bin size used in this calculation is 0.1 km. It can be seen that this high resolution is unnecessary. {\it Top-right:} The plot of collision term associated with emission used in the simulation. {\it Bottom-left:} The plot of the collision term associated with the absorption term used in the simulation. {\it Bottom-right:} The plot of the collision term associated with the direction-changing term used in the simulation.
}
\label{vaconly}
\end{figure}

Using this simple case, it is easy to understand why the length scale of the problem is much larger than $c\mu^{-1}$. The form of the QKEs is such that there isn't a single emitting surface, but an emitting region, which is the same as the neutrino decoupling region. Neutrinos emitted from each radius in the decoupling region oscillates with the length scale given by $c \omega_{\mathrm{vac}}^{-1}$. However, the flavor content at a given location is determined by neutrinos coming from different points in the decoupling region which are not in phase with each other due to varying propagation lengths. The result is a smearing of the oscillatory pattern. This is not something new and has been known for a very long time; when the size of the neutrino emitting region is much larger than the neutrino oscillation length, the oscillatory pattern gets smeared out~\cite{Shalgar:2022rjj}~\footnote{Ref.~\cite{Shalgar:2022rjj} was not the first paper to propose this explanation, but it has been known for so long that I do not know the original source.}.

In Fig.~\ref{vaconly}, I present a calculation in which I first solve eqs~\ref{eom1} and \ref{eom2} without the flavor evolution, including only the advective term and the collision terms. The collision terms corresponding to the emission, absorption, and direction-changing terms are shown in the top-right, bottom-left, and bottom-right panels, respectively. The collision terms are assumed to have no energy dependence and the term $\mathcal{C}_{\textrm{ani}}$ is assumed to be zero. This is the same as the Boltzmann equations. Then I further evolved the system with $\theta_{\textrm{V}} = \pi/4$ and $c\omega_{\textrm{vac}}^{-1}= 0.1$~km. The top-left panel shows the difference in the difference number density due to flavor evolution. One can clearly see that there is no structure of 0.1 km size and that a spatial resolution of 0.1 km is completely unnecessary. In fact, I have checked this explicitly (the result is not shown here). A similar computation involving self-interactions shows the same results as has been demonstrated in Ref.~\cite{Shalgar:2022rjj}.

This can be easily checked by the readers as it is not a very difficult calculation from a numerical point of view. One does not need to use the specific collision terms I have used; any reasonable radial profiles of the collision terms can be used for this simple calculation. Also, since the system is linear one does not even have to evolve the system with time. The equations of motion in the absence of the self-interaction Hamiltonian can be expressed as a set of linear equations that can be solved by LU decomposition or something similar. However, the computational time required to do the calculation this way does not scale very well with the number of radial bins.

Most readers must have realized by now that the misconception that the length scale is the same as $c\mu^{-1}$ perhaps arises from the overreliance on the neutrino-bulb model to gain insights into the phenomenology of collective neutrino oscillations.  
However, I think it will be helpful if another argument is presented to counter the myth regarding the length scale of the problem.

Let us consider the eqs.~\ref{eom1} and \ref{eom2} and see how numerical techniques actually solve them. For the sake of simplicity of understanding let us assume that we are solving eqs.~\ref{eom1} and \ref{eom2} using explicit Euler method. This would be extremely inefficient if we tried but it will convey the argument more simply. The explicit Euler method solves the eqs.~\ref{eom1} and \ref{eom2} through successive application of temporal evolution of the following form:
\begin{eqnarray}
\rho(t + \Delta t) &=& \rho(t) + \Delta t \left(-\vec{v}\cdot\vec{\nabla}\rho(t)-i[H(t),\rho(t)]+\mathcal{C}\right)\\
\bar{\rho}(t + \Delta t) &=& \bar{\rho}(t) + \Delta t \left(-\vec{v}\cdot\vec{\nabla}\bar{\rho}(t)-i[\bar{H}(t),\bar{\rho}(t)]+\bar{\mathcal{C}}\right)\ .
\end{eqnarray}
Here, I have suppressed all the arguments other than time for brevity. The terms $\vec{v}\cdot\vec{\nabla}\rho(t)$ and $\vec{v}\cdot\vec{\nabla}\bar{\rho}(t)$ can be calculated using the central difference method. 
The time step size, $\Delta t$ should be very small for convergence, which I shall discuss in more detail later. However, it should be noted that in the case of QKEs relevant to core-collapse supernovae, the commutator term, $-i[H(t),\rho(t)]$, is always much larger than the advective term, $\vec{v}\cdot\vec{\nabla}\rho(t)$ (the same is true for the case of antineutrinos). Typically, in numerical simulations, one can see that the advective term is much smaller than the self-interaction term. In fact, the numerical error that is inevitably present in any method can be comparable if not larger than the advective term.
This may seem strange indeed as we know that the advective term is very important for solving the QKEs.

It is important to understand the different nature of these two terms. The self-interaction term represented by the commutator leads to a rapid evolution of flavor with respect to time for any given $\cos\theta$ and $r$, irrespective of the presence of the advective term. Consequently, over several hundred time steps the advective term will not see the instantaneous gradient but a gradient in the flavor states that have been averaged over several oscillation periods. It should be pointed out that this averaging is not something that is done by hand but happens due to the nature in which explicit solvers evolve discretized partial differential equations. 
This is true in any explicit method of solving the QKEs as long as sufficiently small time step size is used. 
The same thing should happen if implicit solvers are used with sufficiently small time step size but that defeats the purpose of using implicit solvers. This is one of the many reasons why implicit solvers are not suitable for solving QKEs.
This is a crucial ingredient in understanding the length scale of the problem, which is not easy to grasp or appreciate for someone who has not performed simulations with various numerical techniques and physical setups.

Before I conclude, I would also like to comment on the choice of $\Delta t$, which can be chosen using an adaptive step size algorithm, and depends on the type of solver. Many studies I have seen in the literature use a fixed time step dictated by the Courant condition~\cite{wikicourant}. In that case, one needs to be careful. If one assumes that $c\mu^{-1}$ is the length scale of the problem and that the Courant condition dictates that a Courant factor of 0.5 (or something of that order) will give convergence, then that is not correct. In that case, one is essentially using a time step that is half of the typical time scale of the problem. One can easily see the problem with that by attempting to numerically solve a simple harmonic oscillator of the form $\ddot{x}(t) + \omega^{2} x(t)=0$ with a time-step of $\frac{1}{2\omega}$; the results will be wrong. There are several papers that use this technique of assuming that the length scale of the problem is $c\mu^{-1}$ and applying the Courant condition will give the choice of time step size~\footnote{I have chosen not to cite those papers here for obvious reasons.}. A cursory look at the Wikipedia page on the topic shows that the Courant condition is based on the movement of waves, and we can use the Courant condition to determine the size of the time step if the advective term is the dominant part of time evolution~\cite{wikicourant}. In the case of QKEs, of course, the Courant condition has to be satisfied as far as the time step size is concerned, but the required time step size is typically much smaller than that obtained by setting the Courant factor to $\mathcal{O}(0.1)$.

\section{Conclusions} \label{sec:conclusions}
\label{conclusion}
In this paper, I have attempted the task of dispelling some myths regarding the length scale associated with collective neutrino oscillations. However, this is not an easy task for me because, as with all myths, I do not know the supposed reasons and rationale why people believe in them. I have tried to guess the possible reasons and address them. 

Even for a system with period boundary conditions, it is difficult to justify the need for spatial bin size that is as small as $c\mu^{-1}$. It is only some very specific toy models that one can see a cascade to very small length scales.

I have also discussed physical systems that do not have periodic boundary conditions. Using QKEs without self-interaction term and vacuum term only it is easy to show that the length scale of the system is not related to the frequency of neutrino oscillations. I also explain why, in general, a bin size much larger than $c\mu^{-1}$ is sufficient for numerical convergence.

I have argued that the spatial resolution required in numerical simulations need not be of the order of $c\mu^{-1}$. That leaves the important question unanswered: What should be the spatial resolution used in the numerical simulations of QKEs? The way it is determined in all numerical simulations by physicists -- by performing two simulations with spatial resolutions differing by a factor of two. If the results are the same within the required accuracy that means the results have converged. Any additional assumptions or preconceptions beyond this are unnecessary.

\acknowledgments
Although this paper will not be submitted to any journal, I would appreciate comments, criticisms, and suggestions to improve the document.

\bibliographystyle{JHEP}
\bibliography{lengthscales.bib}
\end{document}